\documentclass[a4paper,twocolumn,fleqn]{article}

\usepackage{amsmath}             
\usepackage{txfonts}               
 \usepackage{mathptmx}            
\usepackage{graphicx}
\usepackage{pfr}                   
\usepackage{stfloats}
\usepackage{siunitx}
\usepackage{hyperref}
\hypersetup{
    colorlinks=true,
    linkcolor=blue,
    filecolor=magenta,      
    urlcolor=blue,
}
\begin{document}

\title{Systematic comparison of VMEC and HINT equilibrium calculations for finite-$\beta$ LHD plasmas}


\author{Albert~CIVIT-BERTRAN\sup{1,2}, Yasuhiro~SUZUKI\sup{1} and Shimpei~FUTATANI\sup{3,4,2}}

\affiliation{
  \sup{1}Graduate School of Advanced Science and Engineering, Hiroshima University, 739-8527 Higashi-Hiroshima, Japan. \\
  \sup{2}Universitat Politècnica de Catalunya - BarcelonaTech (UPC), Barcelona, Spain. \\
  \sup{3}National Institute for Quantum Science and Technology, Naka Institute for Fusion Science and Technology, Naka, Japan. \\
  \sup{4}National Institute for Quantum Science and Technology, Rokkasho Institute for Fusion Energy, Rokkasho, Japan}


\email{albert.civit.bertran@upc.edu}

\begin{abstract}
A systematic comparison between VMEC and HINT equilibrium calculations has been carried out for Large Helical Device plasmas to clarify the influence of the assumption of the nested flux surfaces at finite beta.
Three vacuum magnetic-axis configurations, $R_{\rm axV} = \SI{3.53}{\, m}$, $\SI{3.60}{\, m}$, $\SI{3.85}{\, m}$, are examined for the beta values on the axis in the range $\beta_0 \in [0.0\%, 5.0\%]$. 
The magnetic-axis position, the rotational transform on the axis, and the plasma volume enclosed by the last closed flux surface are compared between the two codes. 
At low-$\beta_0$, VMEC and HINT give consistent equilibria, indicating that the nested flux surfaces are largely preserved. 
Above a configuration-dependent critical $\beta_0$, however, the two solutions begin to diverge, indicating that the nested flux surfaces assumption is compromised. 
In HINT, the enclosed plasma volume decreases at higher beta because the stochastic magnetic field evolves near the plasma edge, whereas VMEC cannot represent this flux surface breaking due to its assumption of nested flux surfaces. 
These results show that the 3D equilibrium responses in LHD equilibria become increasingly important from inward- to outward-shifted configurations, mainly through Pfirsch-Schlüter current-driven perturbations of the magnetic field and the resulting edge stochasticity.
\end{abstract}



\maketitle  


In stellarators, the 3D magnetohydrodynamic (MHD) equilibrium has been widely computed using the VMEC \cite{Hirshman_1983, Hirshman_1986} solver, which assumes perfectly nested flux surfaces to solve the energy principle. However, in stellarators, the increase in plasma pressure alters the equilibrium magnetic field topology by introducing magnetic islands and stochastic regions, compromising the validity of the assumption of nested flux surfaces. In zero net-current devices, the magnetic islands are driven by the Pfirsch-Schlüter (PS) current \cite{PS_current}. The breaking of the magnetic field lines by the overlapping of magnetic islands shrinks the plasma volume enclosed by the last closed flux surface (LCFS), which has been observed to limit the equilibrium beta more severely than the Shafranov shift \cite{Hayashi_1990, Suzuki_2020a}, where beta is $\beta = 2\mu_0  p / B^2$, where $p$ and $B$ are the plasma pressure and magnetic field, respectively. The HINT code \cite{Suzuki_2006, Suzuki_2017} is a 3D MHD equilibrium code that solves the dynamic equations for the magnetic field and pressure using a relaxation method, without assuming nested flux surfaces. HINT can resolve the existence of magnetic islands and stochastic regions in finite-$\beta$ equilibria in stellarators \cite{Suzuki_2020a, Suzuki_2016, Suzuki_2020b} and tokamaks \cite{Suzuki_2017}. In this work, a systematic comparison of VMEC and HINT calculations has been performed. This study was conducted on Large Helical Device (LHD) plasmas. The LHD device vacuum  magnetic axis can be positioned at $R_{\rm axV} =\SI{3.50}-\SI{3.90}{\, m}$ with an average minor radius $a \sim \SI{0.6}{\, m}$ and axis magnetic field $B_0 \leq \SI{3.0}{\,T}$. The configuration $R_{\rm axV} =\SI{3.6}{\, m}$ is referred to as the `standard' configuration, $R_{\rm axV} <  \SI{3.6}{\, m}$ as the `inward-shifted' configuration, and $R_{\rm axV} > \SI{3.6}{\, m}$ as the `outward-shifted' configuration. The device has $M=10$ toroidal field periods and $L=2$ poloidal winding number. The helical pitch $\gamma = 1.254$ and the canceling rate of the quadrupole field $B_q=100\%$ have been used.  In this study, the finite-$\beta$ effects have been compared for on-axis beta values $\beta_0 \in[0.0\%, 5.0\%]$ in increments of $0.5\%$, where $\beta_0 = 2p_0\mu_0/B_0^2$, with $p_0$ and $B_0$ the pressure and magnetic field strength on the magnetic axis, respectively. The initial pressure profile in VMEC and HINT has been defined as $p = p_0(1-s)$, where $s$ is the normalized toroidal flux.
\begin{figure}[t]
  \centering
  \includegraphics[width=\linewidth]{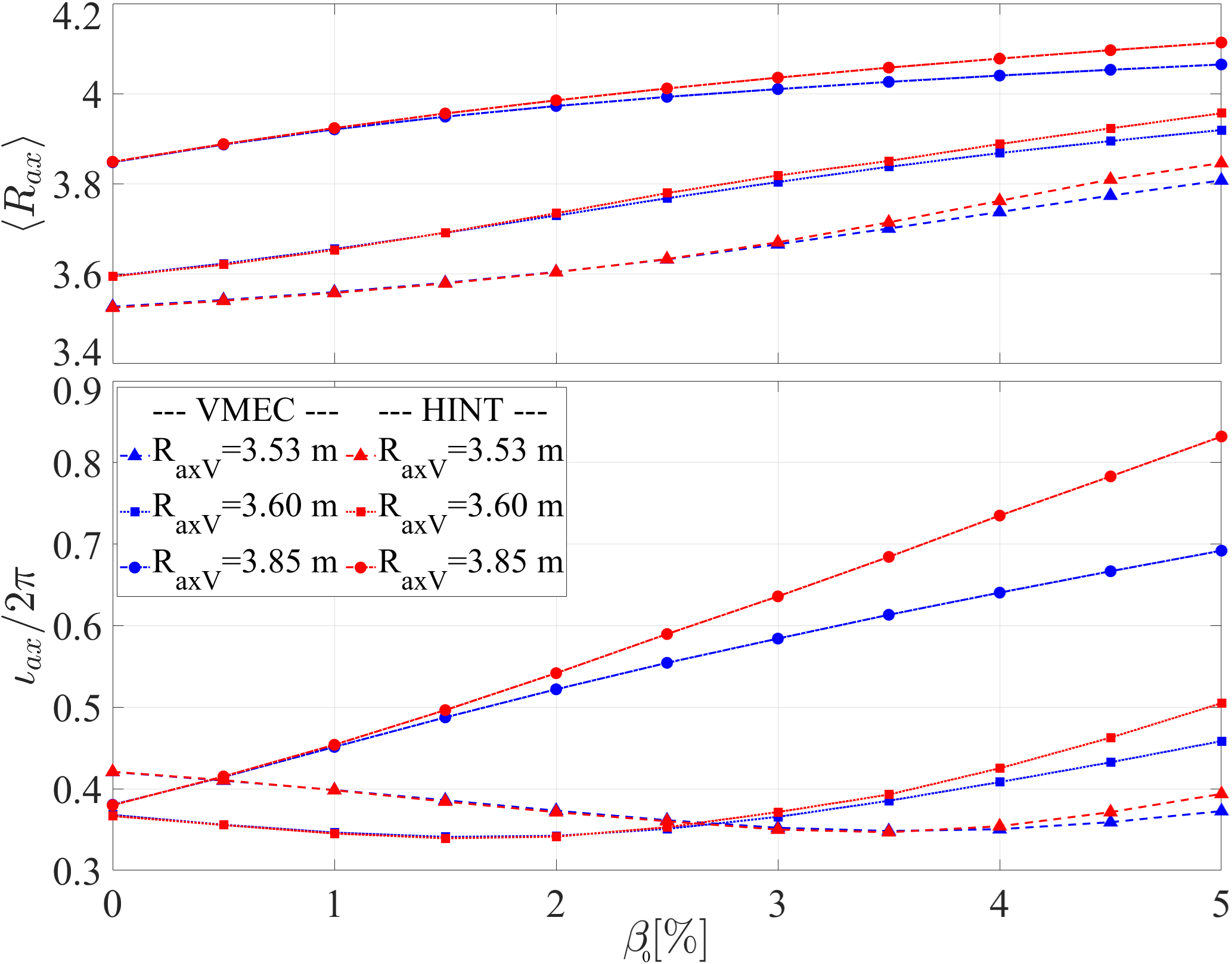}
  \caption{Top) $\langle Rax\rangle$ position and bottom) $\iota/2\pi$ on magnetic axis for different vacuum magnetic axis configurations $R_{\rm axV} = \SI{3.53}{\, m}$, $\SI{3.60}{\, m}$, $\SI{3.85}{\, m}$ against $\beta_0$.}
  \vspace{-15pt}
  \label{fig:RaxV_iota_vs_beta}
\end{figure}
\begin{figure}[t]
  \centering
  \includegraphics[width=\linewidth]{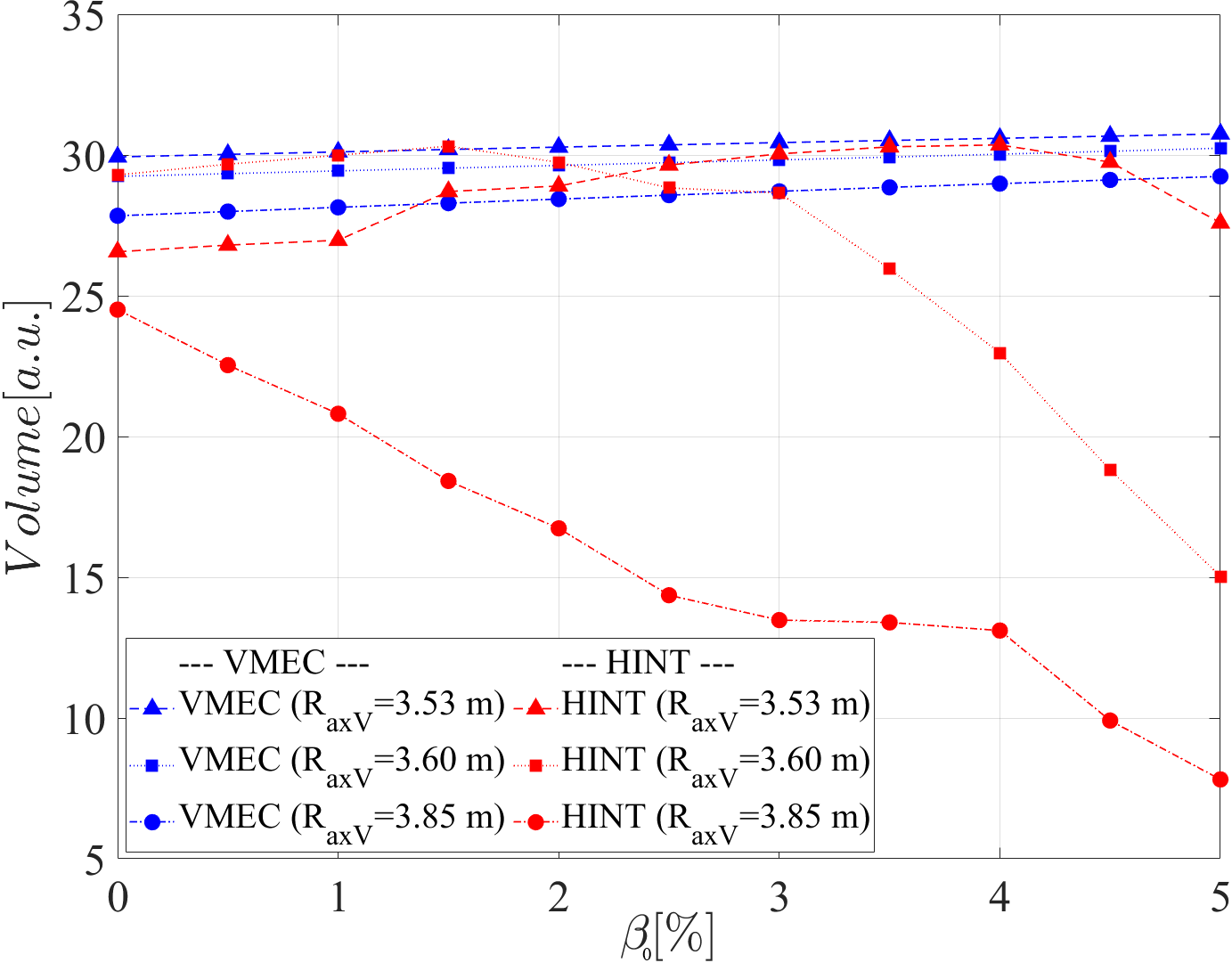}
  \caption{Volume enclosed by the clear LCFS for vacuum configurations $R_{\rm axV} = \SI{3.53}{\, m}$, $\SI{3.60}{\, m}$, $\SI{3.85}{\, m}$ against $\beta_0$.}
  \vspace{-22pt}
  \label{fig:Volume}
\end{figure}

Figure~\ref{fig:RaxV_iota_vs_beta} shows top) the averaged magnetic axis position $\langle R_{\rm axV}\rangle$, and bottom) the axis rotational transform $\iota/2\pi$  as $\beta_0$ increases. For the studied LHD configurations, there exists a critical $\beta_0$ value after which HINT resolves a larger Shafranov shift and rotational transform than VMEC does. The stochastization of the edge magnetic field lines reduces the magnetic pressure resulting from flux-surface compression, allowing the plasma to shift further outwards than the constraint imposed by nested flux surfaces, thereby sustaining higher magnetic pressure. The critical $\beta_0$ value depends on the vacuum configuration, decreasing from inward- to outward-shifted configurations. To characterize the breaking of the edge flux surfaces, the volume enclosed by the clear LCFS has been studied. Figure~\ref{fig:Volume} compares the volume for the different $\beta_0$ values. For inward-shifted and standard configurations, the plasma volume for low-$\beta_0$ increases with $\beta_0$ in both VMEC and HINT calculations. However, after the critical $\beta_0$ value, the volume enclosed by the LCFS in HINT code tends to decrease for the three $R_{\rm axV}$  configurations. The volume difference indicates that the 3D equilibrium responses~\cite{Suzuki_2017} shrink the HINT-LCFS by rendering the magnetic field stochastic at the edge. In the absence of net toroidal current, the parallel component of the plasma equilibrium current, the PS current, is generated by $\nabla\cdot\Vec{j}=0$. When the PS current resonates with the magnetic field, resonant magnetic fields are evolved. The radial width of the resonant magnetic field grows with the parallel current and decreases with the local magnetic shear. The PS current is proportional to the pressure gradient and to the toroidicity of the configuration. The nested flux surfaces break when the resonant magnetic fields grow sufficiently to form magnetic islands, and neighboring magnetic islands overlap each other. The difference in the critical $\beta_0$ value for the three $R_{\rm axV}$ configurations arises from differences in the magnetic properties of the vacuum configuration. The helical ripple of the LHD configuration grows with $R_{\rm axV}$ \cite{Yamada_2011}, increasing the PS current in the plasma. At the same time, the magnetic shear of the outward-shifted configuration decreases compared to the standard and inward-shifted configurations. The two phenomena cause the 3D equilibrium responses to be stronger in the outward-shifted configuration, where islands grow larger at lower beta and stochastically perturb the magnetic field at overlap. 

HINT code resolves magnetic islands and stochastization of the magnetic field that occur in stellarators, where the assumption of nested flux surfaces is compromised as a result of the 3D equilibrium response. The systematic comparison between VMEC and HINT codes proves that for low-$\beta_0$ configurations, the 3D MHD equilibria of the two codes are comparable, with nested flux surfaces preserved. However, as $\beta_0$ increases, HINT calculates a larger Shafranov shift and larger rotational transform on the magnetic axis due to the 3D equilibrium responses. The plasma volume enclosed by the clear LCFS in HINT code shrinks with increasing $\beta_0$ due to the breaking of the flux surfaces, in comparison to the volume of VMEC which monotonically increases with $\beta_0$. The critical $\beta_0$ value after which the nested closed flux surfaces become significantly compromised depends on the vacuum configuration, where stronger toroidicity and lower magnetic shear make the configuration more sensitive to the 3D equilibrium responses.

The resulting publication has the support of the Joan Oró predoctoral grants program AGAUR-FI (2023 FI-3 00065) from the Secretary of Universities and Research of the Research and Universities Department of the Government of Catalonia, and from the European Social Fund Plus. This publication is based upon work supported by “Beques Santander – Ayudas predoctorales 2025”. This work received funding from the Spanish Ministry of Science (Grant No. PID2020-116822RBI00). The authors gratefully acknowledge RES computational resources provided by BSC in MareNostrum to RES-FI-2025-03-0067. 


\vspace{-10pt}

\begin{thebibliography}{9}
\bibitem{Hirshman_1983}
  S.P. Hirshman, J. C. Whitson 1983 \textit{Phys. Fluids} \textbf{26} (12): 3553–3568.
\bibitem{Hirshman_1986}
  S.P. Hirshman, W.I. van Rij and P. Merkel 1986 \textit{Comput. Phys. commun.} \textbf{43} 143
\bibitem{PS_current}
  D. Pfirsch and A. Schlüter 1962, MPI/PA/7/62, (Max-Plank Institut für Physik und Astrophysik, München).
\bibitem{Hayashi_1990}
  T. Hayashi \textit{et al}., 1990 Phys. Fluids B \textbf{2}, 329.
\bibitem{Suzuki_2020a}
  Y. Suzuki \textit{et al}. 2020 \textit{Phys. Plasmas} \textbf{27} 102502
\bibitem{Suzuki_2006}
  Y. Suzuki, \textit{et al}., 2006 \textit{Nucl. Fusion} \textbf{46} L19
\bibitem{Suzuki_2017}
  Y. Suzuki 2017 \textit{Plasma Phys. Control. Fusion} \textbf{59} 054008
\bibitem{Suzuki_2016}
  Y Suzuki and J Geiger 2016 \textit{Plasma Phys. Control. Fusion} \textbf{58} 064004
\bibitem{Suzuki_2020b}
  Yasuhiro Suzuki 2020 \textit{Plasma Phys. Control. Fusion} \textbf{62} 104001
\bibitem{Yamada_2011}
  H. Yamada for the LHD Experiment Group 2011 \textit{Nucl. Fusion} \textbf{51} 094021
\end{thebibliography}
\end{document}